\documentclass[aps,twocolumn,amsmath,amssymb,preprintnumbers]{revtex4}
\usepackage{amsmath} \usepackage{amsfonts} \usepackage{amssymb}
\usepackage{bbm}
\usepackage{epsfig}
\usepackage{graphics}
\usepackage{graphicx}
\newcommand{\be}{\begin{equation}}
\newcommand{\ee}{\end{equation}}
\newcommand{\ba}{\begin{eqnarray}}
\newcommand{\ea}{\end{eqnarray}}
\newcommand{\nn}{\nonumber}

\usepackage{todonotes}

\begin{document}

\title[ ]{Probabilistic cellular automaton for quantum particle in a potential}

\author{C. Wetterich}
\affiliation{Institut  f\"ur Theoretische Physik\\
Universit\"at Heidelberg\\
Philosophenweg 16, D-69120 Heidelberg}

\begin{abstract}
We propose that a quantum particle in a potential in one space dimension can be described by a probabilistic cellular automaton. While the simple updating rule of the automaton is deterministic, the probabilistic description is introduced by a probability distribution over initial conditions. The proposed automaton involves right- and left-movers, jumping from one cell to a neighboring one. They change their direction of motion at each randomly distributed disorder or scattering point. The continuum limit of an infinite number of cells yields a Dirac equation, and in the non-relativistic limit the familiar Schrödinger equation, with potential determined by the spacetime-distribution of scattering points. These equations describe the time evolution of the probabilistic information for the position of the particle. All quantum rules for observables, both for discrete possible measurement values and continuous expectation values, follow from the classical statistical laws.
\end{abstract}

\maketitle

We all have learned quantum mechanics by exercises solving the Schrödinger equation for a particle in a potential in one space dimension, finding the typical quantum phenomena of tunneling through a potential barrier, the reflection and transmission laws or bound states in an harmonic potential. Possible measurement values are typically discrete, as occupation numbers $n_{p}(x)= 0 , 1 $ for a particle being present or not in a given space interval, or cell, or ``detector", located at $x$. The quantum laws yield probabilities for finding one of the two possible values or, equivalently, the expectation value $\langle n_{p}(x) \rangle$. The occupation numbers $n_{p}(x)$ can be identified with bits of an automaton that take the same discrete values. In this note we propose an updating rule for the automaton such that in the continuum limit of a very large number of cells the expectation values  $\langle n_{p}(x) \rangle$, as well as the expectation values for other observables, follow precisely the same time evolution as for quantum mechanics of a particle in a potential. This continuum limit requires a probabilistic setting for the automaton, as specified by a probability distribution for initial bit configurations.

A particle present at the position $x$ corresponds to a bit $n_{p}(x)$ taking the value one. For one possible realization a one-particle state can be seen as a bit configuration with all bits except one taking the value zero. For the occupation number observables the possible measurement values are the values of the bits for different positions or cells $x$.
One-particle states are highly correlated since the measurement $n_{p}(x)=1$ at some point $x$ predicts $n_{p}(y)=0$ for all $y \neq x$. For massless particles without a potential we consider two sorts of bits. Right-movers move at every discrete time step from $t$ to $t+\varepsilon$ one position to the right, i.e. from $x$ to $x+\varepsilon$, while for left-movers the motion is in the opposite direction from $x$ to $x-\varepsilon$. If there are several species of bits a particle is present at $x$ if one of these bits takes the value one. Instead of a single non-zero bit we can also describe one-particle states as excitations of a half-filled vacuum. The location of the particle (or hole) is then the position of the additional non-zero bit (or missing non-zero bit).

We next place on the $(t,x)$-lattice disorder points at which particles change their direction of motion, switching between right-movers and left-movers.
 The intuitive picture of our automaton is simple: At each time step a particle continues to move or is reflected, according to a disorder point absent or present at is actual position. For every time $t$ one has a different fixed distribution of disorder points. On the trajectory of a particle with given initial position reflections occur in certain irregular time intervals. The random space-time distribution of disorder points is chosen with weights such that in the average the probability to find the particle at time $t$ at position $x$ is the same as for quantum mechanics. The quantum mechanical potential is given by the distribution of disorder events, and vice versa. Our setting has the structure of a cellular automaton~\cite{NEU,ULA,ZUS,WOL,PREDU,TOF}, where the updating in each cell is determined by the neighboring cells. It does not repeat the same updating at every time step, however. Due to the random distribution of disorder points we may call it a ``random cellular automaton".

One may visualize our setting as a generalized two-dimensional Ising model with fixed disorder. The disorder points have a fixed distribution on a two-dimensional lattice. One of the lattice directions is associated with time.  The occupation numbers at given space-time points $(t,x)$ are identified with Ising spins $s(t,x)=2n_{p}(t,x)-1$. The interactions between the Ising spins are chosen such that only the time-sequences of bit configurations which are allowed by the updating rule of the automaton have non-zero probability. The probabilistic aspect enters only by a probability distribution for the Ising spins at the boundary of the lattice at ``initial time". This system defines an overall positive probability distribution or partition function or functional integral for ``events" in space time, i.e. particles being present or absent, or Ising spins being up or down, at lattice points $(t,x)$. In general, this system can describe a type of quantum field theory with an arbitrary number of fermionic particles. We focus here on one-particle states, given by an appropriate initial condition for which the probability is zero if more than one particle is present. 

If this classical statistical system can describe the time evolution of all expectation values for quantum mechanics for a particle in a potential, this simple example may shed new light on an old debate if quantum mechanics can emerge from classical statistics. This debate seemed to have found a negative answer by no-go theorems~\cite{BELL,CHSH}. It was argued, however, that these theorems do not apply if simultaneous classical probabilities for two observables either do not exist or are not available for a subsystem (``incomplete statistics")~\cite{CW1}. For many systems, including Dirac fermions in an electromagnetic potential in four dimensions~\cite{CW2} or a quantum particle in a potential in three dimensions~\cite{CW3}, an explicit evolution law for a classical probability distribution has been found which reproduces the quantum time evolution of expectation values. Furthermore, for every classical statistical system described by an overall partition function or functional integral for all times a quantum formalism can be based on classical wave functions or density matrices encoding the time-local probabilistic information on suitable time-hypersurfaces~\cite{CWIT,CWQF}. In this case the evolution equation for the time-local probabilistic information is no longer ad hoc, following now from the functional integral in close analogy to Feynman's path integral approach to quantum mechanics~\cite{FEYN}.

%Absatz
The question arises if suitable overall classical probability distributions can lead to the same evolution laws for the density matrix as for interesting quantum systems. Probabilistic cellular automata are interesting candidates in this direction. The automaton property guarantees the characteristic quantum feature that no information is lost during the evolution. The probabilistic initial conditions can be incorporated in an overall classical statistical partition function or functional integral. The simple evolution law of the automaton allows for a practical realization of the time-evolving probabilities. 

The probabilistic formulation distinguishes our setting from the interesting approach of a deterministic description of quantum mechanics by cellular automata~\cite{HOOFT,ELZ,HO1,HO2,HO3}. We will see that the probabilistic setting is crucial for the continuum limit which is a key aspect of the present work.
Probabilistic automata equivalent to two-dimensional fermionic quantum field theories with interactions have been constructed explicitly~\cite{CAF1,FPCA}, and extended to four dimensions~\cite{CW4D}. The present note aims for a particularly simple example which may allow for explicit numerical simulations of the automaton. This may constitute a further convincing piece that it is indeed possible to embed quantum mechanics in classical statistics. 

The Schrödinger equation for a particle in a potential $V(\chi)$,
\begin{equation}
\label{A}
i\partial_{t}\chi (t,x)=-\frac{1}{2m}\partial_{x}^{2}\chi(t,x) +V(x)\chi(t,x)\ ,
\end{equation}
involves the complex wave function $\chi(t,x)$. (We use units $\hbar =c=1$). For a numerical solution it may be discretized with discrete time steps $t=m_{t}\varepsilon$ and unitary step evolution operator $U(t)$,
 \begin{align}
 \label{B}
 &\chi(t+\varepsilon,x)=\sum_{y} U(t; x,y)\chi(t,y)\;,\quad U=\exp(-i\varepsilon H)\ ,\nn\\
 &H(x,y)=\bigl{(}-\frac{1}{2m}\partial_{x}^{2}+V(x)\bigr{)}\delta_{x,y}\ .
  \end{align}
  Here also the space coordinate is discrete, with $\partial_{x}^{2}$ $\delta_{x,y}$ involving lattice derivatives typically connecting neighboring $x$ and $y$. 
  The expectation values of the particle numbers $n_{p}(t,x)$ are given in this discrete setting by
  \begin{equation}
  \label{C}
   \langle n_{p}(t,x)\rangle =|\chi(t,x)|^{2}\;\; ,\; \quad\sum_{x}\langle n_{p}(t,x)\rangle =1\ .
  \end{equation}
They are predicted by the solution of the Schrödinger equation with given initial condition $\chi(t_{\textup{in}},x)$. It is our aim to reproduce these expectation values by a random cellular automaton with a suitable distribution of disorder events representing a given potential $V(x)$. The crucial additional ingredient as compared to a simple ``deterministic" solution of a discretized Schrödinger equation is the direct realization of the discrete observables and the associated probabilities or expectation values. Our proposal for a probabilistic automaton can be seen as a ``physical realization" of a quantum particle in a potential.

We will obtain the Schrödinger equation as the non-relativistic limit of the evolution of a Dirac fermion. Dirac fermions in two dimensions consist of four Majorana-Weyl fermions. We will therefore consider four species of bits $n_{R,\eta}(t,x)$, $n_{L,\eta}(t,x)$ at every space-time point. The index $R$, $L$ indicates right-movers and left-movers. The index $\eta=\pm 1$ can be associated with particles and holes with respect to a half-filled vacuum state. In this case the one-particle state has precisely one particle more or less than the half filled state. Alternatively, we may associate $\eta$ with two ``colors". A deterministic one particle state has at a given $t$ precisely one of the $\langle n_{\gamma ,\eta}(t,x)\rangle$ equal to one, and all others zero ($\gamma =R, L$). A probabilistic automaton specifies for each possible one particle bit-configuration a probability $p_{\gamma,\eta}(t,x)$. The updating rule of the automaton specifies how $p(t+ \varepsilon)$ is computed from $p(t)$: at $t+\varepsilon$ the probability for a configuration is precisely equal to the one from which it has originated at $t$. 

It is convenient to express this setting in terms of a real classical wave function $q_{\gamma ,\eta}(t,x)$ which is the square root of the probability distribution
\begin{equation}
\label{D}
p_{\gamma ,\eta}(t,x)=q_{\gamma ,\eta}^{2}(t,x)\ .
\end{equation}
By normalization of the probability distribution the wave function at any given $t$ is a real unit vector 
\begin{equation}
\label{E}
\sum_{\gamma ,\eta , x} q_{\gamma ,\eta}^{2}(t,x)=1 \ .
\end{equation}
The evolution of the wave function is given by a real orthogonal step evolution operator $\widehat{S}$
\begin{equation}
\label{F}
q_{\gamma ,\eta}(t+\varepsilon ,x)=\sum_{\gamma' , \eta' , y}\widehat{S}(t;x,y)_{\gamma\eta , \gamma' \eta'} q_{\gamma' ,  \eta'}(t,y)\ .
\end{equation}
For an automaton the step evolution operator is a unique jump matrix with precisely one element $\pm 1$ in each row and column, and zeros otherwise. This guarantees that each one-particle bit configuration ($\gamma', \eta', y$) is updated to precisely one new bit configuration $(\gamma , \eta, x)$ according to the corresponding non-zero element of $\widehat{S}$. 

We assume that there is no difference in the updating rule for particles and holes (or the two colors), such that $\widehat{S}_{\gamma\eta , \gamma'\eta'}=\widehat{S}_{\gamma\gamma'}\delta_{\eta\eta'}$.
The remaining step evolution operator $\widehat{S}_{\gamma\gamma'}(t; x,y)$ is a 2$\times$2 matrix field depending for a given $t$ on $x$ and $y$.
We introduce a complex structure and a complex two-component wave function by~
\begin{equation}
\label{G}
\varphi_{\gamma}(t,x)=\frac{1+i}{\sqrt{2}}q_{\gamma +}(t,x)+\frac{1-i}{\sqrt2}q_{\gamma -}(t,x) \ .
\end{equation}
Complex conjugation maps particles to holes and vice versa. This is the setting how the complex two-component single particle wave function for a Dirac spinor consists of two Weyl spinors $\varphi_{R}$, $\varphi_{L}$, which each are composed of two real Majorana-Weyl spinors. Since the step evolution operator is real one has in the complex formulation 
\begin{equation}
\label{H}
\varphi=\begin{pmatrix}
\varphi_{R}\\ \varphi_{L}
\end{pmatrix}\; ,\quad \varphi(t+\varepsilon , x)=\sum_{y} \widehat{S}(t; x,y)\varphi(t,y)\ .
\end{equation}
It will be our task to find unique jump step evolution operators for which eq.~\eqref{H} describes the propagation of a Dirac fermion in a potential. The non-relativistic limit of the Schrödinger equation will then be straightforward.

%Absatz
Our starting point is the step evolution operator
\begin{equation}\label{1}
\widehat{S}(t)=\widehat{S}_{f}\widehat{S}_{V}(t)\ ,
\end{equation}
where the free part,
\begin{equation}\label{2}
(\widehat{S}_{f})(x,y)=\begin{pmatrix}
\delta_{x,\, y+\varepsilon} &, &0\phantom{\Big{|}} \\
0&,&\ \delta_{x,\, y-\varepsilon} 
\end{pmatrix}\ ,
\end{equation}
moves the right-movers one position to the right and the left-movers one position to the left. In our discrete setting neighboring positions are separed by $\varepsilon$. In order to avoid redundancy we restrict to a sublattice with $x=2m_{x}\varepsilon$ for even $m_{t}$ and $x=(2m_{x}+1)\varepsilon$ for odd $m_{t}$, $m_{x}$ integer.
We assume at every given $t=m_{t}\varepsilon$ a certain number of disorder points or scattering events at positions $\overline{x}_{i}(t)$. The scattering part of the step evolution operator $\widehat{S}_{V}$ switches at each scattering point a right-mover to a left-mover and vice-versa. We can write this diagonal operator in exponential form
\begin{equation}\label{3}
\widehat{S}_{V}(t; x, y)=\exp\lbrace -i\varepsilon H_{V}(t,x)\rbrace\delta_{x,y}\ ,
\end{equation}
with
\begin{equation}\label{4}
H_{V}(t,x)=\frac{\pi\tau_{2}}{2\varepsilon}\sum_{i}\delta_{x,\overline{x}_{i}(t)} \ .
\end{equation}
Indeed, for every position $x=\overline{x}_{i}(t)$ the scattering operator produces a factor $-i\tau_{2}$ which switches right- and left-movers, while for all other positions it is unity. The operator \eqref{1} is a unique jump matrix.

The free part can be written as well as an exponential of a matrix $H_{f}(x)$
\begin{equation}\label{5}
\widehat{S}_{f}(x,y)=\exp(-i\varepsilon H_{f})(x,y)\,
\end{equation}
with
\begin{equation}\label{6}
H_{f}(x,y)=\sum_{p}D^{-1}(x,p)p D (p,y)\tau_{3}=\tilde{P}(x,y)\tau_{3}\ .
\end{equation}
The discrete momentum $p$ is periodic in $\pi/\varepsilon$, 
\begin{equation}
\label{7}
p=\frac{\pi q}{\varepsilon M_{x}}\quad , \quad |q|\leqslant \frac{M_{x}}{2}\ ,
\end{equation}
with integer $q$. Here $M_{x}$ is the total number of $x$ points at a given $t$, taken for simplicity on a circle with lengh $L=2M_{x}\varepsilon$,
and $M_{x}+q$ is identified with $q$. With the discrete Fourier transform
\begin{equation}
\label{8}
D^{-1}(x,p)=\frac{1}{\sqrt{M_{x}}}\exp(ipx)=D^{*}(p,x)\ ,
\end{equation}
we recognize $\tilde{P}(x,y)$ as a discrete momentum operator in the position basis. 

The $M_{x}\times M_{x}$ matrix $D$ is unitary, $\sum_{p}D^{*}(x,p)D(p,y)=\delta_{x,y}$, such that multiplication with $D$ corresponds to a change of basis for the wave function.
The identity of eq.~\eqref{2} and eq.~\eqref{5} is proven by transforming $\widehat{S}_{f}$ to momentum space. Due to the factor $\tau_{3}$ the right- and left-movers are displaced in opposite directions.  The possibility of a change of basis as the Fourier transform is one of the great advantages of the use of wave functions for encoding the probabilistic information. It cannot be realized for probability distributions.

%Absatz
Let us now consider a large number $N_{t}$ of evolution steps $(\Delta t = N_{t}\varepsilon )$
\begin{align}\label{9}
\varphi(t+\Delta t)&=\prod_{n_{t}=0}^{N_{t}-1}\Big{[}\exp(-i\varepsilon H_{f})\exp \bigl( -i\varepsilon H_{V} (t+n_{s}\varepsilon)\bigr)\Big{]}\varphi (t)\nn\\
&=\exp(-i\Delta t\overline{H})\varphi (t)=\exp (-i\varepsilon N_{t}\overline{H})\varphi (t)\ .
\end{align}
The matrix $H_{f}$ does not commute with the diagonal matrix $H_{V}$, and the order in the product is with larger $n_{t}$ to the left. (In momentum space $H_{f}$ is diagonal and $H_{V}$ has off-diagonal elements.) Since every single evolution step is unitary also the sequence of $N_{t}$ steps is unitary and eq.~\eqref{9} defines a hermitian $M_{x}\times M_{x}\, $-matrix $\overline{H}$.
The operator $\overline{H}$ involves averaging over time and will be identified with an effective Hamiltonian. We will later also proceed to space averaging.

We may write $\overline{H}=H_{0}+\Delta H$, where $\Delta H$ arises from commutators of $H_{f}$ and $H_{V}$ and is formally of the order $\varepsilon$. For $H_{0}$ one finds
\begin{equation}
\label{10}
H_{0}=H_{f}+\frac{1}{N_{t}}\sum_{n_{t}=0}^{N_{t}-1} H_{V}(t+n_{t}{\varepsilon})=H_{f}+\overline{V}(x)\tau_{2}\ .
\end{equation}
(Our notation often omits unit matrices in position space.)
The quantity $\overline{V}(x)$ involves the time averaged number of scattering events $\overline{n}(x)$ at $x$ within the given time interval,
\begin{equation}
\label{11}
\overline{V}(x)=\frac{\pi}{2\varepsilon}\overline{n}(x)\; ,\quad \overline{n}(x)=\frac{1}{N_{t}}\sum_{n_{t}=0}^{N_{t}-1} \sum_{i}\delta_{x,\overline{x}_{i}(t+n_{t}\varepsilon)}\ ,
\end{equation}
or the corresponding number density $n(x)=\overline{n}(x)/(2\varepsilon)$, $\overline{V}(x)=\pi n(x)$. We may split $\overline{V}(x)$ into a homogeneous part $m$ and an inhomogeneous part $V(x)$,
\begin{equation}
\label{12}
\overline{V}(x)=m+V(x)\ .
\end{equation}
The quantity $m$ is identified with the mass of the particle and $V(x)$ with the potential. 

%Absatz
Our next step defines a continuum limit for which $\Delta H$ can be neglected. For this purpose we keep the time interval $\Delta t$ fixed and increase the numer $N_{t}$ of time points in this interval, $N_{t}\rightarrow \infty$, $\varepsilon\rightarrow 0$. For the corresponding sequence of automata with decreasing $\varepsilon$ the number $\widehat{n}(x) $ of scattering events within the time interval~$\Delta t$ at a given position $x$ is kept fixed, resulting in fixed $n(x)$ and $V(x)$ due to $\overline{n}(x)=\widehat{n}(x)/N_{t}\sim \varepsilon$, 
\begin{equation}
\label{13}
n(x)=\frac{\overline{n}(x)}{2\varepsilon}=\frac{\widehat{n}(x)}{2\varepsilon N_{t}}=\frac{\widehat{n}(x)}{2\Delta t}\; ,\,\quad \overline{V}(x)=\frac{\pi \,\widehat{n}(x)}{2\Delta t}\ .
\end{equation}
We further assume wave functions that remain smooth in the continuum limit $\varepsilon\rightarrow 0$. In this case the relevant momenta in Fourier space are much smaller than $\pi/\varepsilon$ ($|q|\ll M_{x})$. For $\varepsilon \rightarrow 0$ the periodicity of momenta becomes unimportant, expressing the momentum operator by a lattice derivative, $\tilde{P}=-i\widehat{\partial}_{x}$. In turn, for smooth wave functions this becomes the partial derivative,
\begin{equation}
\label{14}
H_{f}=-i\partial_{x}\tau_{3}\ .
\end{equation}
We can switch to continuous time and space. For time independent $\overline{H}$ in eq.~\eqref{9} the wave function $\varphi(t)$ obeys then the differential evolution equation
\begin{equation}
\label{15}
i\partial_{t}\varphi=\overline{H}\varphi=\Big{[}-i\partial_{x}\tau_{3}+\bigl{(}m+V(x)\bigr{)}\tau_{2}\Big{]}\varphi\ .
\end{equation}

Eq.~\eqref{15} is the two-dimensional Dirac equation with a potential $V(x)$. Using the Dirac matrices $\gamma^{0}=-i\tau_{2}$, $\gamma^{1}=\tau_{1}$, $\partial_{0}=\partial_{t}$, $\partial_{1}=\partial_{x}$ it can be written in the form
\begin{equation}\label{16}
\Big{[}\gamma^{\mu}\partial_{\mu}+m+V(x)\Big{]}\varphi=0\ .
\end{equation}
For $V(x)=0$ this equation is invariant under Lorentz-transformations. It describes the propagation of a free fermion with mass $m$ and its antiparticle. The Schrödinger equation obtains in the non-relativistic limit $|p|\ll m$ provided that $|V(x)|\ll m$.  Indeed, for 
\begin{equation}
\label{16A}
\varphi=\frac{1}{\sqrt{2}}e^{-imt}\begin{pmatrix}
\chi-i\partial_{x}\chi/(2m)\\  i\chi-\partial_{x}\chi/(2m)
\end{pmatrix}\ ,
\end{equation}
the function $\chi$ obeys the Schrödinger equation~\eqref{A}. There is a second solution of this type with $\varphi\sim e^{imt}$, whose complex conjugate describes the one-particle state for the antiparticle.

For a sufficiently smooth distribution of disorder events the neglected part $\Delta H$ of the Hamiltonian $\overline{H}$ is of the order $\varepsilon [\tilde{P}\tau_{3},\, \overline{V}(x)\tau_{2}]$.
As compared to the kinetic part $\tilde{P}\tau_{3}$ of $H_{0}$ this is suppressed by a factor $\varepsilon m$, while as compared to the piece $\overline{V}(x) \tau_{2}$ in the $H_{0}$ the suppression factor is $\varepsilon \tilde{P}$.
For the continuum limit $\varepsilon\rightarrow 0$ we can omit $\Delta H$. The argument is parallel to the treatment of the non-commuting kinetic and potential pieces in the Hamiltonian for the derivation of the Feynman path integral. Here it is based on the Baker-Campbell-Haussdorff formula for products of exponentials of matrices. For more details see ref.~\cite{FPCA}.

The continuum limit should also involve an averaging or coarse graining in space. This is best done by a folding of the Dirac equation~\eqref{15},\eqref{16}, $\overline{\varphi}(x)=\int \textup{d}y f(x-y)\varphi(y)$, with $f(x)$ having support in a certain interval $\Delta x$ around zero and normalized such that the normalization of the averaged wave function $\overline{\varphi}$ is preserved. The elimination of the high-momentum components of the wave function by the averaging plays no role if $\varphi$ is already sufficiently smooth. The potential $V(x)$ and the number of disorder points $\widehat{n}(x)$ in the interval $\Delta t$ get replaced by space-averaged values $\langle V(x)\rangle$ and $\langle \widehat{n}(x)\rangle$. The result is a smoothening of these functions, and the possibility to realize $\langle\widehat{n}(x)\rangle\ll 1$ by a sparse distribution of disorder events in space. This allows us to obtain a continuum description even for time intervals $\Delta t$ much smaller than $m^{-1}$, since for $\overline{V}(x) \approx m$ eq.~\eqref{13} becomes
\begin{equation}
\label{17}
m\Delta t  =\frac{\pi}{2}\langle \widehat{n}\rangle =\frac{\pi\, n_{\textup{int}}}{2\,N_{x}}\  ,
\end{equation}
where $n_{\textup{int}}$ is the total number of scattering events in the combined interval $\Delta t\Delta x$ and $N_{x}$ the number of sites in the interval $\Delta x$. The large numbers $n_{\textup{int}}$ needed for a smooth distribution can be compensated by even much larger $N_{x}=\Delta x/\varepsilon$ for $\varepsilon\rightarrow 0$ . 

We conclude that the probabilistic random cellular automaton is indeed equivalent to quantum mechanics in the continuum limit $\varepsilon\rightarrow 0$. 
The particle mass can be much smaller than the ultraviolet momentum cutoff $\varepsilon^{-1}$ if the fraction of disorder events is sufficiently small, $m\varepsilon\sim n_{\textup{int}}/(N_{t}N_{x})\ll 1$. It is an interesting question if reasonable finite systems can be found which allow for a numerical test of the equivalence with quantum mechanics. As an example, let us consider for $x$ a circle with length $2\cdot 10^{3}\varepsilon$, $M_{x}=10^{3}$, and a total time extension $M_{t}=10^{4}$. We consider intervals with $N_{t}=N_{x}=100$ and place in the interval $\Delta t\Delta x$ in the average 100 disorder events, leading to $\varepsilon m =\pi/200$, $\Delta tm=\pi/2$. Labelling by $i$ the $\Delta t $- intervals and by $j$ the $\Delta x$-intervals the numbers $n_{\textup{int}}(i,j)$ should not depend on $i$ in order to maintain time-translation invariance. 
A dependence of $n_{\textup{int}}(i,j)$ on $j$ induces a potential $V(x)$. For each $i$ we may randomly place only 95 disorder events in the intervals $j= 1, 2, 3, 4, 6, 7, 8, 9$, while in the intervals $j= 5, 10$ one locates 120 disorder events. The corresponding static potential is given by $V(x)/m=-0.05$ for $0<x<800\varepsilon$ and $1000\varepsilon<x<1800\varepsilon$, with potential barriers $V(x)/m=0.2$ for $800\varepsilon<x<1000\varepsilon$ and $1800\varepsilon<x<2000\varepsilon$.

One can start at $t=0$ with an initial wave function (and associated probability distribution) which has support only in the intervals $j= 1, 2, 3, 4$, such that the probability to find the particle initially in the intervals $j=\,$5-10 vanishes.
For a small enough kinetic energy one expects tunneling through the potential barrier towards the region $j=6,7,8,9$. This can be investigated quantitatively by evolving the cellular automaton in time. One needs to follow 4000 trajectories, one for each particle initially located at $x$ and of type $(\gamma, \eta)$. Right-movers move to the right and left-movers to the left until they encounter a disorder event and change direction. As long as only probabilities are needed for an observable one simply assigns to each point on a trajectory the same probability as for its initial point. 
For every $t$ this leads to a probability distribution for the $10^{3}$ space points. All observables that are functions of occupation numbers - this includes correlations - can be directly computed and compared with the results of a solution of the quantum mechanical Schrödinger equation with the same initial wave function for $\chi$. 

Quantum mechanics admits further observables as momentum. For the cellular automaton the expectation value of the momentum can be computed by switching the momentum operator $ \tilde{P}$ defined by eq.~\eqref{6} between the wave functions, $\langle P\rangle =q^{T}\tilde{P}q$, in complete analogy to quantum mechanics. The computation of this or similar expectation values requires to follow the signs of the elements of the real wave function in addition to the magnitude. Momentum is a ``statistical observable" which measures properties of the probabilistic information, as periodicity~\cite{FPCA}.
In this sense it has a similar conceptual status as temperature. The momentum observable has no fixed value for a particular bit-configuration. Simultaneous probabilities and classical correlation functions for momentum and occupation numbers are not defined. This is the simple reason why Bell's inequalities~\cite{BELL,CHSH} cannot be applied.

Our setting can be generalized directly to states with more than one particle. An arbitrary bit configuration at a given time corresponds to a configuration of fermionic occupation numbers. We can identify the general wave function with a multi-fermion wave function for a fermionic many body system in the occupation number basis. This is mapped to a discrete real formulation of a quantum system. The probabilistic automata can be equivalent to a discretized fermionic quantum field theory~\cite{CAF1}.

The formulation of certain discretized fermionic quantum field theories with interactions as probabilistic cellular automata sheds light on the possible origin of the random distribution of disorder events that we have employed in the present note. In a fermionic quantum field theory the scattering occurs due to interactions, and this is the same for the corresponding automata. Mass terms and a potential for single particle states can be induced by expectation values of fermionic bilinears or higher composites. Mass generation by chiral symmetry breaking is indeed a common feature of many fermionic quantum field theories~\cite{NJL}. From the point of view of single particle trajectories the expectation values of fermion composites are indeed seen as random scattering events.

In conclusion, we have demonstrated that the simple quantum system of a particle in a potential in one space dimension can be described by a classical statistical system. Numerical simulation of the corresponding random probabilistic automaton seems feasible. Extensions to more than one space dimension need further work. If random  probabilistic cellular automata can be found for this case as well, this could explain why the propagation of particle-like objects in certain macroscopic classical systems shows a behavior close to quantum mechanics~\cite{COU,EDD}.

%\newpage

%\bibliography{refs}

\end{document}